\documentclass[aps,pre,preprint,a4paper,onecolumn]{revtex4}
\usepackage[]{amsmath}
\usepackage{amssymb}
\usepackage[dvips]{graphicx}
\usepackage{color}
\usepackage{bm}
\usepackage{tabularx}
\usepackage{mathtools}
\usepackage{algpseudocode}
\usepackage{enumitem}
\usepackage{multirow}
\usepackage{epstopdf}
\usepackage{ulem}

\newcommand{\rom}[1]{\uppercase\expandafter{\romannumeral #1\relax}}

\begin{document}
\title{DP-GEN: A concurrent learning platform for the generation of reliable deep learning based potential energy models}
\author{Yuzhi Zhang}
\affiliation{Beijing Institute of Big Data Research, Beijing 100871, People's Republic of China}
\affiliation{Yuanpei College of Peking University, Beijing 100871, People's Republic of China}
\author{Haidi Wang}
\affiliation{School of Electronic Science and Applied Physics, Hefei University of Technology, Hefei 230601, People's Republic of China}
\author{Weijie Chen}
\affiliation{Academy for Advanced Interdisciplinary Studies, Peking University, Beijing 100871, People's Republic of China}
\author{Jinzhe Zeng}
\affiliation{School of Chemistry and Molecular Engineering, East China Normal University, Shanghai 200062, People's Republic of China}
\author{Linfeng Zhang}
\email{linfengz@princeton.edu}
\affiliation{Program in Applied and Computational Mathematics, Princeton University, Princeton, NJ, USA}
\author{Han Wang}
\email{wang_han@iapcm.ac.cn}
\affiliation{Laboratory of Computational Physics, Institute of Applied Physics and Computational Mathematics, Huayuan Road 6, Beijing 100088, People's Republic of China}
\author{Weinan E}
\email{weinan@math.princeton.edu}
\affiliation{Program in Applied and Computational Mathematics, Princeton University, Princeton, NJ, USA}
\affiliation{Beijing Institute of Big Data Research, Beijing 100871, People's Republic of China}

\date{\today}

\begin{abstract}
In recent years, promising deep learning based interatomic potential energy surface (PES) models have been proposed that can potentially
allow us to perform molecular dynamics simulations for large scale systems with quantum accuracy.
However, making these models truly reliable and practically useful is still a very non-trivial task. A key component  in this task is the generation
of datasets used in model training.
In this paper, we introduce the Deep Potential GENerator (DP-GEN), an open-source software platform that implements the recently proposed "on-the-fly" learning procedure [Phys. Rev. Materials 3, 023804]  and is capable of generating uniformly accurate deep learning based PES models in a way that minimizes human intervention and the computational cost for data generation and model training. 
DP-GEN automatically and iteratively performs three steps: exploration,  labeling, and training. 
It supports various popular packages for these three steps: LAMMPS for exploration, Quantum Espresso, VASP, CP2K, etc. for labeling, and DeePMD-kit for training.
It also allows automatic job submission and result collection on different types of machines, such as high performance clusters and cloud machines, and is adaptive to different job management tools, including  Slurm, PBS, and LSF.
As a concrete example, we illustrate the details of the process for generating a general-purpose PES model for Cu using DP-GEN.
\end{abstract}

\maketitle
\section{Introduction}
In recent years, machine learning (ML) has emerged as a promising tool for the field of molecular modeling.
In particular, ML-based models have been proposed to address a long-standing issue, 
the accuracy-{\it vs}-efficiency dilemma when one evaluates the potential energy surface (PES), a function of atomic positions and their chemical species, and its negative gradients with respect to the atomic positions, namely the interatomic forces.
From a first-principles point of view, PES is derived from the many-particle Schr\"odinger equation under the Born-Oppenheimer approximation, and the interatomic forces are given naturally by the Hellman-Feynman theorem. 
To this end, the {\it ab initio} molecular dynamics (AIMD) scheme, wherein accurate PES and interatomic forces are obtained within the density functional theory (DFT) approximation, has been most widely adopted~\cite{kohn1965self,car1985unified,marx2009ab}.
Unfortunately, the cost of AIMD restricts its typical applications to system sizes of hundreds of atoms and the time scale of $\sim$~100~$ps$. 
In the opposite direction,  efficient empirical PES models, or force fields (FF), allow us to perform much larger and longer simulations;
but their accuracy and transferability is often an issue.
ML seems to have the potential to change this situation: a good ML model  trained on {\it ab initio} data should have the efficiency of FF models while maintaining {\it ab initio} accuracy.

Developing ML-based PES models involves two components,  data generation and model construction.
To date, most discussions have focused on the second component. 
Two important issues are: A good functional form (e.g. kernel based models or neural networks) and respecting physical
constraints of the PES, such as the extensiveness and  symmetry properties.
In this regard, two representative classes of models have emerged:
the kernel-based models like the Gaussian Approximation Potential~\cite{bartok2010gaussian} and the neural network (DNN) based models like the Behler-Parrinello model~\cite{behler2007generalized} and the Deep Potential  model~\cite{han2017deep,zhang2018deep}. 
In particular, the smooth version of the Deep Potential model is an end-to-end model that satisfies the requirements mentioned above~\cite{zhang2018end}.

There have also been some efforts on open-source software along this line~\cite{artrith2016implementation,wang2018kit,scḧutt2018schnetpack,yao2018tensormol}.
Of particular relevance to this work is the DeePMD-kit package~\cite{wang2018kit}, which has been developed to minimize the effort required to build DNN-based PES models and to perform DNN-based MD simulation. 
DeePMD-kit is interfaced with TensorFlow~\cite{tensorflow2015}, one of the most popular deep learning frameworks, making the training process highly automatic and efficient.
DeePMD-kit is also interfaced with popular MD packages, such as the LAMMPS package~\cite{plimpton1995lammps} for classical MD and the i-PI package~\cite{Ceriotti2014iPI} for path integral MD.
Thus, once one has a good sets of data, there are now effective tools for training 
Deep Potentials that can be readily used to perform efficient molecular dynamics simulation for all different kinds of  purposes.

In comparison, much less effort has gone into the first component mentioned above: data generation. 
In spite of the tremendous interest and activity, very few have taken the effort 
to make sure that the dataset used to train the ML-based PES is truly representative enough.
Indeed data generation is often quite {\it ad hoc}.
Some notable exceptions are found in~\cite{podryabinkin2017active,smith2018less,zhang2019active}. 
In Ref.~\cite{smith2018less}, an active learning procedure was proposed based on an existing
unlabled dataset. 
Some data points in that set are selected to be labled, and the result is then used to train the ML model. 
The procedure ensures that the selected dataset is at least representative of the original unlabeled dataset.
In Refs.~\cite{podryabinkin2017active,zhang2019active}, one begins with no data,  labeled or
unlabeled,  and explores the configuration spaces following some systematic procedure.  
For each of the configurations encountered, a decision is made as to whether that configuration should be labeled.
The exploration procedure is designed to ensure that the unlabeled dataset, i.e. all the configurations encountered in the exploration, 
is representative of all the situations that the ML-based model is intended for.
Even though these procedures were also described as ``active learning'', there is a difference
since in these procedures, one does not have the unlabeled data to begin with and choosing
the unlabeled data judiciously is also an important part of the whole procedure.

To highlight this difference, we will call the procedures in Refs.~\cite{podryabinkin2017active,zhang2019active} ``concurrent learning''. 
By ``concurrent learning'', we mean that one does not have any data to begin with, labeled or unlabeled, and the data is generated on the fly as the training proceeds. 
The generation of the data and the learning is an interactive process to  ensure that one obtains an ``optimal'' dataset which is on one hand representative enough and on the other hand as small as possible.
This is in contrast to  ``sequential learning'' in which the data is generated beforehand and the training of the model is performed afterwards.
It also differs from active learning in the sense that active learning starts with unlabeled data. 
In a way the purpose of active learning is to find the smallest dataset that needs to be labeled in an existing unlabeled dataset.

The actual concurrent learning procedure goes as follows.
One uses  different sampling techniques (such as direct MD at different thermodynamic conditions, 
enhanced sampling, Monte Carlo) based on the current approximation of the PES to explore
the configuration space. 
An efficient error indicator (this is the error in the PES) is then used to monitor the snapshots generated during the sampling process.
Those that have significant errors will then be selected and sent to a labeling procedure, in which accurate \textit{ab initio} energies and forces are calculated and added to the training dataset.
A new approximation of the PES is obtained by training with the accumulated training dataset.
These steps are repeated until convergence is achieved, i.e., the configuration space has been explored sufficiently, and a representative set of data points has been accurately labeled. 
At the end of this procedure, a uniformly accurate PES model is generated.
We refer to Ref.~\cite{zhang2019active} for more details.

In order to carry out such a procedure efficiently, one also needs reasonable computational resources. 
Since many tasks can be done in parallel, one needs to implement automatic and efficient parallel processing algorithms.
Taking the exploration stage for example, it may happen that dozens to hundreds of MD simulations are executed simultaneously with different initial configurations under different thermodynamic conditions.
If these tasks are executed manually, it will require a great deal of human labor, not to mention the compromise in efficiency.
It would be even worse if one wants to utilize different computational resources in different concurrent learning  steps, e.g., a high performance cluster(HPC) with most advanced GPU nodes for training and an HPC with a vast number of CPU nodes for labeling. 
Selecting a machine with the most available computational power among a group of candidate machines is also an issue. 
For all these reasons, we feel that it would be useful for the molecular and materials simulation community to have an open-source implementation of the concurrent learning  procedure which, 
among other things, can automatically schedule the iterative process, dispatch different computational tasks to different computational resources, and collect and analyze the results.

In this paper, we introduce DP-GEN, an open-source concurrent learning platform and 
software package for the generation of reliable deep learning based PES models, in a way that minimizes the computational cost and human intervention.
We describe the implementation of DP-GEN, which is  based on the procedure proposed in Ref.~\cite{zhang2019active}. 
We will focus on two modules, the scheduler and the task dispatcher. 
A modularized coding structure for the scheduler is designed, making it possible to incorporate different methods or software packages for the three different components in the
concurrent learning procedure:  exploration, labeling, and training. 
The dispatcher module is prepared for handling a huge number of tasks in a high-throughput 
fashion, and it is made compatible with different kinds of machines and popular job 
scheduling systems.  

This paper is organized as follows. 
In Section II we present the basic methodology that the DP-GEN workflow follows.
In Section III we introduce the details of the software, including how the concurrent
 learning process is scheduled and how different tasks are dispatched.
In Section IV, we give a concrete example, in which a general purpose Deep Potential
 model for Cu is generated using DP-GEN.
Conclusions and outlooks are given in the last Section.

\section{Methodology}\label{sec:method}
The DP-GEN workflow contains a series of successive iterations.
Each iteration is composed of three steps: exploration, labeling, and training.
We denote by $E_\omega(\mathcal{R})$, abbreviated $E_\omega$, the PES represented by the DP model, where $\mathcal{R}$ denotes atomic positions and $\omega$ denotes the parameters.
An important point throughout the DP-GEN procedure is that we have an ensemble of models $\{E_{\omega_1}, E_{\omega_2}, \dots, E_{\omega_\alpha}, \dots\}$ trained from the same set of data but with difference in the initialization of model parameters  ${\omega_\alpha}$. 
${\omega_\alpha}$ evolves during the training process, which is designed to minimize 
the loss function, a measure of the error between DP and DFT results for the energies, 
the forces, and/or the virial tensors.
Since the loss function is a non-convex function of ${\omega}$, such a difference of initialization leads to different minimizers after training, and therefore different PES models. 
Around configurations where there is enough training data, the different PES models should all be quite accurate and therefore produce predictions that are close to each
other. 
Otherwise one would expect that the predictions from the different models will scatter with a considerable variance.
Therefore, the variance of the ensemble of predictions from the ensemble of models for a particular configuration can be used as an error indicator and criterion for whether the configuration should be selected for labeling.

{\it Exploration}: This has two components: an efficient sampler and an error indicator.
The goal of the sampler is to efficiently explore the configuration space.
Assume that we have an initial configuration denoted by $\mathcal R_0$. 
The sampler, in general, is a propagator that evolves an initial configuration of a system to a series of new configurations,
\begin{align}\label{eqn:sampler}
 \mathcal R_t = \varphi_t (\mathcal R_0, E_{{\omega}})
\end{align}
where $\varphi_t$ can be either deterministic or stochastic,
with $t$ labeling a continuous, or discrete, series of operations,
and $E_{{\omega}}$ indicates that the DP model is parameterized by ${\omega}$.
Available implementations of the sampler include direct MD, MD with enhanced sampling techniques, the Markov chain Monte Carlo (MCMC) approach, and the genetic algorithm (GA), etc.
Both the sampler and the initial configurations should be chosen to ensure that all configurations of practical interest are approximately visited with sufficiently high frequency.
It is worth noting that since the sampler ~\eqref{eqn:sampler} uses the DP model - rather than an {\it ab initio} routine - to evaluate potential energies and atomic forces, the exploration is significantly more efficient than AIMD simulations.

\begin{figure}[htbp]
\includegraphics[width=8cm ]{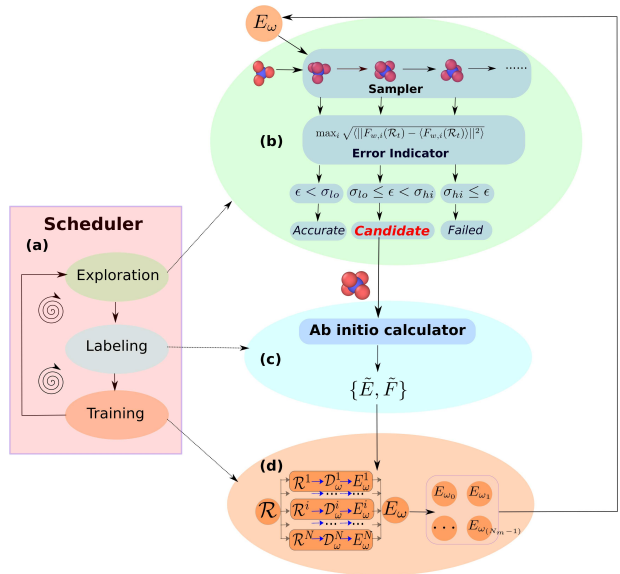}
\centering
\caption{Schematic illustration of the DP-GEN scheme. (a) The scheduler iteratively and automatically promotes exploration, labeling and training steps. 
(b) An ensemble of DP models $E_\omega$ is obtained in the training process.  
(c) An exploration strategy based on DPMD
is taken as an example. Given fixed structures as starting points, $E_{\omega}$ is used to drive MD simulations and sample a series of configurations. 
For each configuration, the error indicator $\epsilon$, defined by the max force deviation of atomic forces predicted by the DP model ensemble, is calculated. 
Only those satisfying the criterion $\sigma_{lo}\leq\epsilon<\sigma_{hi}$ are selected as candidates for labeling.
(d) In the labeling step, the {\it ab initio} calculator computes first-principles  energies $\tilde{E}$ and forces  $\tilde{F}$ for the candidates.
}
\label{fig:scheme}
\end{figure}

Next, given a configuration $\mathcal R_t$, we define the error indicator $\epsilon_t$ as the maximal standard deviation of the atomic force predicted by the model ensemble, i.e.,
\begin{align}
 &\epsilon_t = \max_i\sqrt{\langle\Vert F_{w,i}(\mathcal R_t)-\langle F_{w,i}(\mathcal R_t)\rangle\Vert^2\rangle},
 \\ \nonumber
 &F_{w,i} (\mathcal R_t) = -\nabla_i E_{w}(\mathcal R_t)\label{eq: model_devi}
\end{align}
where $F_{w,i}$ denotes the predicted force on the atom with index $i$, and $\nabla_i$ denotes the derivative with respect to the coordinate of the $i$-th atom.
The ensemble average $\langle ... \rangle$ is taken over the ensemble of models, and is estimated by the average of model predictions, i.e.
\begin{align}
 \langle\nabla_i E_{w}(\mathcal R_t)\rangle \approx \frac{1}{N_m} \sum_{\alpha=1}^{N_m} \nabla_i E_{w_\alpha}(\mathcal R_t)
\end{align}
Due to the way the error indicator is defined, we also call it the model deviation. 
The reason for using the predicted forces, rather than energies, to evaluate the 
model deviation is that the force is an atomic property and is sensitive to the local 
accuracy, especially when a failure happens.
Energy is a global quantity and does not seem to provide sufficient resolution.
As shown in Fig.~\ref{fig:scheme}, given an upper and lower bound of the trust levels, $\sigma_{hi}$ and $\sigma_{lo}$ , those structures whose model deviations $\epsilon$ fall between the bounds will be considered candidates for labeling. 
Thus, the candidate set of configurations is defined by
\begin{align}
 \{ \mathcal R_{n\Delta t} | {n\in I_\textrm{cand}} , \quad I_{\textrm{cand}} = \{\, n\, | \sigma_{lo} \leq \epsilon_{n\Delta t} < \sigma_{hi} \}\}
 \end{align}
DP-GEN will then randomly select user-defined number of configurations from the candidate set, and pass them into the next step.

To better illustrate the idea, we take the copper system as an example, for which the user targets at a uniform accuracy over the thermodynamic range of temperature from 0~K to $2T_m$ and pressure from 1~Bar to 50,000~Bar, where $T_m = 1357.77$~K denotes the melting point at the ambient condition.
The exploration strategy can be designed as follows. 
We divide the temperature range $[50\mathrm{K}, 2T_m)$ equally into four intervals, i.e.~$[50\mathrm{K}, 0.5T_m)$, $[0.5T_m, T_m)$, $[T_m, 1.5T_m)$, $[1.5T_m, 2T_m)$, and explore them successively. 
In each temperature interval we run 8 iterations with increasing number of MD simulations and increasing length of trajectories. 
For example in the first iteration, 600 MD simulations of 2~ps are performed, while in the last iteration the number of MD simulation increases to 2400 and the length trajectory increases to 6~ps.
In each iteration, 5 temperatures conditions that evenly divide the temperature interval and 8 pressure conditions, i.e. 
  1,
  10
  100,
  1,000,
  5,000,
  10,000,
  20,000 and
 50,000~Bar, 
are explored simultaneously by Isothermal-isobaric (NPT) DP-based MD simulations.
The initial configurations of these MD simulations are prepared by randomly perturbing fully relaxed standard crystal structures, including face-centered cubic (fcc), hexagonal-closed-packed
(hcp), and body-centered cubic (bcc) structures. 
The configurations along MD trajectories are recorded at a time interval of $\Delta t = 0.02$~ps, and
those with model deviation  $0.05\textrm{eV/\AA} \leq \epsilon_{n\Delta t} < 0.20\textrm{eV/\AA}$ are selected and passed to the labeling stage.

{\it Labeling}: This step  calculates the reference {\it ab initio} energies $\tilde{E}$ and forces $\tilde{F}$ for the selected configurations $\mathcal{R}_{n\Delta t}$ from the {\it exploration} step. 
The process can be done by first-principles-based schemes, including quantum chemistry, quantum Monte Carlo, and DFT, etc.  
The results are called labels.
These labels are then added into the training dataset, to be used later for retraining.

{\it Training}: 
We adopt an advanced version of the Deep Potential (DP) method proposed in Ref.~\cite{zhang2018end}. 
The DP model considers $E_\omega$ as a sum of contributions from atomic local environments $\mathcal{R}^i$, i.e. , $E_\omega = \sum_i E_\omega^i(\mathcal{R}^i)$.
$\mathcal{R}^i$ contains the coordinates of $i$'s neighboring atoms within a $r_c$ cutoff radius, 
and it is mapped, through an embedding network, onto a so-called feature matrix $\mathcal{D}_\omega^i$. 
Such a construction guarantees the invariance of the PES under translation, rotation, and 
permutation among identical particles.
$\mathcal{D}_\omega^i$ is then mapped, through a fitting network, to $E_\omega^i$.
Since ${\omega}$ is composed of the parameters in the embedding network and the fitting network, we also call it the network parameters.

\section{Software}
\subsection{Overview}
Implemented with Python, DP-GEN provides a user-friendly interface.
The master process can be started via a single line of command: 
\begin{verbatim}
    dpgen run PARAM MACHINE
\end{verbatim}
where the arguments PARAM and MACHINE are both the names of parameter files in the json format that specify the user's demands. 

DP-GEN is composed of two major modules.  
First, DP-GEN serves as a scheduler, which follows the aforementioned concurrent learning scheme and generates computational tasks iteratively for the three steps: exploration, labeling and training.
Second, DP-GEN serves as a task dispatcher, which receives tasks from the scheduler and automatically submits tasks to available computational resources, collects results when a task finishes, and sends the results back to the scheduler.

\begin{figure}[]
\centering
\includegraphics[width = 8cm]{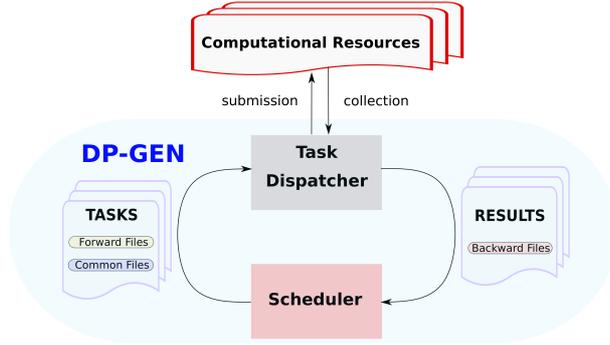}
\caption{(Color online) Overview of the structure of the DP-GEN package, which has two major modules, scheduler and task dispatcher. The scheduler prepares and sends the calculation tasks to the task dispatcher. By communicating with computational resources, 
the task dispatcher automatically handles job submission and the collection of the results.
Results are then sent back to the scheduler, and the master process of DP-GEN enters the next step.
}
\label{fig:flow}
\end{figure}

Fig. \ref{fig:flow} shows the communication between the scheduler module and dispatcher module. 
The scheduler prepares and sends the calculation tasks to the task dispatcher, and in return, it receives results collected by the dispatcher. 
By communicating with the computational resources and using available job management tools,
the task dispatcher automatically handles job submission and result collection.
Typically, a task is composed of two kinds of files, forward files and common files.
Forward files are specific for different tasks, while common files contain universal settings for all tasks. 
The results contained in backward files are then sent back from the dispatcher to the scheduler when the task is finished.

\subsection{DP-GEN scheduler}
The scheduler maintains the information flow between different steps of the concurrent
 learning procedure.
It always works on the machine on which the master process of DP-GEN runs. 
To manipulate format transformations between data files generated by different software, the scheduler utilizes an auxiliary open-source Python package named \texttt{dpdata}, available at GitHub~\cite{dpdata}, developed by the authors.
Details for the implementation of the three steps are as follows. 

{\it Exploration}. At this moment, DP-GEN only supports the use of the LAMMPS package~\cite{plimpton1995lammps}, which has to be interfaced with the DeePMD-kit package to perform DP-based MD (DPMD). 
The scheduler prepares exploration tasks, i.e., the DP models \texttt{graph.00x.pb}, input files \texttt{in.lmp}, and initial structures \texttt{conf.lmp} required by LAMMPS.
In return, it collects the configurations sampled by LAMMPS and selects a subset, according to the criterion for the model deviation, for the labeling step.

The exploration process is based on a sampler, which produces abundant configurations and saves the trajectories in the \texttt{traj} folder with a predefined frequency. 
For each configuration, the model deviation is calculated , and the results are saved in \texttt{model$\_$devi.out}.

The scheduler categorizes all the configurations sampled by the DP model into the following three types. 
DP-GEN will show users the distributions of configurations belonging to different types.
\begin{itemize}
\item $\epsilon_{t} < \sigma_{lo}$. This means that the prediction for the current structure $\mathcal R_{t}$ is accurate within a small tolerance. 
Consequently, there is no need to perform an expensive {\it ab initio} calculation for it. 
\item $\sigma_{hi} \leq \epsilon_t$. 
This means that the model deviation is so large that the prediction on the configuration $\mathcal R_t$ is highly unreliable.
If this happens, the configuration $\mathcal R_t$ is probable to be unphysical as well.
For example, there might be atoms too close to each other. 
This situation often happens in the first several iterations, where the exploration is driven by a relatively poor model. 
Such a configuration should not be labeled and should not be explored in an equilibrium simulation by a more reliable model.
\item $\sigma_{lo} \leq \epsilon_t < \sigma_{hi}$. 
Based on the reasoning above, structures corresponding to this situation are selected as candidates for labeling.
\end{itemize}

{\it Labeling}. In this step, DP-GEN currently supports VASP~\cite{kresse1996efficiency,kresse1996efficient},  Quantum-Espresso~\cite{QE-2017}, Gaussian~\cite{gaussian16}, and CP2K~\cite{Hutter2014cp2k}. 
The scheduler prepares different input files according to the requirements of the 
different software.
In return, it receives first-principles energies, forces, and/or virials, and adds them to the training data.

Selected configurations from the {\it exploration} step will be calculated using one of these software. 
DP-GEN allows and recommends users to take the whole control of their settings for {\it ab initio} calculations. 
The results can be analyzed by the scheduler and new labeled data will be added to the training dataset.
It should be emphasized that the scheduler can handle the case where the expected convergence of the self-consistent-field calculation, adopted by many {\it ab initio} calculations, is not achieved. 
In this case, the results are considered unqualified and thus excluded from the training dataset.

{\it Training}. In this step, the DeePMD-kit package~\cite{wang2018kit} is used for training.
The scheduler prepares training tasks, i.e., the training data and several \texttt{input.json} files required by DeePMD-kit. 
In return, it generates several DP models for efficient exploration in the next step.

The key information provided by the training files \texttt{input.json} are the paths to the training data and the hyper-parameters that define the network structure and the training scheme.
Notice that the training files only differ by the random seeds that initialize the network parameters.
In addition, when the training data are accumulated as DP-GEN proceeds, the scheduler adds the directories of the new data to \texttt{input.json}. 
After training, the resulting DP models, named as \texttt{graph.00x.pb} where~\texttt{x} goes from 0 to $N_m-1$ , are collected by the scheduler.

To keep track of the DP-GEN process, the scheduler takes records of the different steps.
Once the scheduler steps forward, it saves checkpoints in \texttt{record.dpgen}. Therefore, when a fatal failure occurs and user intervention is required, the scheduler can restart from the latest checkpoint automatically. 
The user can also retrieve to any previous step when he/she finds that the current results are problematic and wants to modify the parameters.

\subsection{Task Dispatcher}
Since the number of tasks passed from the scheduler is enormous, 
it would be tedious and time-consuming to manually manage script submission and result collection.
This motivates us to equip DP-GEN with an intelligent task dispatcher.
In general, the dispatcher automatically uploads input files, submits scripts for the calculations,  keeps maintaining the job queues, and collects results when a job finishes.  
The dispatcher shows its strength in handling not only various types of calculation tasks, but also accommodating different kinds of available computational resources. 

\begin{figure}[htbp]
\begin{center}
\includegraphics[width = 8.4cm]{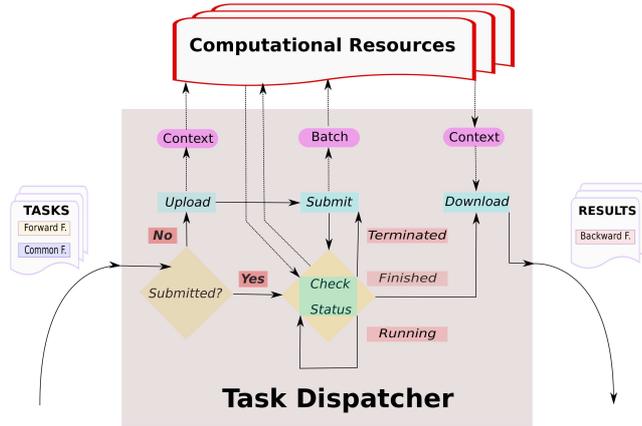}
\caption{(Color online) Schematic illustration of the task dispatcher of DP-GEN. 
The dispatcher receives the tasks to be calculated, and sends the results by communicating with the computational resources.
When a task is sent into the dispatcher, the dispatcher first checks whether it has been submitted. 
If not, it will experience mainly four successive stages in the dispatcher. 
I. Upload. Through the class Context, the dispatcher sends all input files onto machines. 
II. Submit. The class Batch deals with providing adaptive scripts to designate proper computational resources for the task.  Then, a job to calculate the task will be executed on the machine.   
III. Check$\_$status. By querying the machines periodically, the dispatcher checks the status of all jobs, which can be running, terminated or finished. Running status leads to nothing but waiting for the next check. If a job is found terminated, it will be submitted again. Files in jobs with finished status will be downloaded. 
IV. Download. The class Context takes charge of transferring files containing valuable results back into the dispatcher. Till now, all the four stages have been fulfilled, and a  calculation task has finished.
Additionally, those tasks submitted before will enter the third stage check$\_$status directly.
}
\label{fig:dispatcher}
\end{center}
\end{figure}

A typical workflow of the dispatcher is presented in Fig.~\ref{fig:dispatcher} and is composed of following functions:

\textit{Check submission.}
First of all, the dispatcher will check whether the tasks have been submitted or not. 
If not, the dispatcher will upload files and construct the submission script from scratch.  
This will form a queue composed of tasks to be executed. 
Otherwise, the dispatcher will recover the existing queue and collect results from those tasks that have finished, instead of submitting the scripts again. 
Users will benefit a lot from such a design, in the sense that repeated computations are avoided when restarting the DP-GEN master process. 

\textit{File uploading and downloading.} When the DP-GEN master process is running on the login node of an HPC, it shares the file system with computational nodes via a network file system.
File uploading and downloading can be simply implemented by Python modules \texttt{os} and  \texttt{shutil}.  
To enhance the I/O efficiency, instead of copying or removing files, the dispatcher operates on the symbolic links or directly moves files. 
When the DP-GEN master process is running on a machine other than the machines that perform
the actual computations, files are transmitted via \texttt{ssh}. 
The DP-GEN provides a uniform interface for these two situations, so file transmission is adapted to the connection type and can be invoked easily. 
Moreover, new protocols for file transmission can be implemented easily.

\textit{Job submission.} After uploading forward and common files onto the computational machines, DP-GEN generates scripts for executing the desired computational tasks. 
The job scripts are adaptive to different kinds of machines,  including workstations, high performance clusters with a job scheduling system, and cloud machines:
\begin{itemize}
\item \textit{Workstation}.  This allows users to run DP-GEN on a single computational machine, such as a personal laptop. In this situation, DP-GEN prepares shell scripts which can be directly executed.
\item \textit{HPCs with a job scheduling system}, such as Slurm~\cite{slurm}, PBS~\cite{pbs}, and LSF~{\cite{lsf}}.
The resources to execute a job require setting in the submission scripts,  such as the number of CPU/GPUs and maximal execution wall-time, etc.
DP-GEN provides an end-to-end interface for users, and transforms the settings in MACHINE file to the submission scripts, which can be accepted by the job scheduling system.  
\item \textit{Cloud machines}. The script for cloud machines is similar to one for workstation. 
The only difference is that the DP-GEN dispatcher runs additional commands to launch or terminate machines before or after the job execution, respectively.
These commands are compatible with the application programming interface (API) that the cloud machine service provides. 
\end{itemize}

\textit{Job monitoring}. 
Monitoring the status of each job submitted is of vital importance. 
The DP-GEN dispatcher is able to identify the job status and react correspondingly by communicating with all kinds of the machines introduced above. 
If a job is running, the dispatcher will do nothing.
If a job is terminated, the dispatcher will try to resubmit the job up to three times. 
If the job still cannot be successfully executed, it is highly likely that the job settings are problematic, e.g., the parameters might be improper or the configurations might be unphysical.
In this case, user intervention is required.
When a task is finished, the dispatcher downloads the backward files and passes them to the scheduler. 
The task dispatcher accomplishes its mission after all tasks are finished, and then the scheduler will step forward.

At last, it should be emphasized that DP-GEN is fairly automatic, general, and robust. Once the input files are prepared and  DP-GEN runs successfully, there is no need for extra labor. 

\section{Examples}

 \begin{table*}[htbp]
\centering
\caption{\label{tab:TestDesign}Summary and description of the tests used to validate the DP model for Cu. 
}
\begin{tabular}{clp{300pt}}
\hline
order&test&description\\
\hline
0&equilibrium state&
the atomization energy $E_{\text{am}}$ and equilibrium
volume per atom $V_0$ at 0 K\\
1&equation of state& the energy along a given range of volume\\
2&elasticity&  the elastic constants compared with experimental, DFT and MEAM results \\
3&vacancy& the formation energy of vacancy defect compared with experiment, DFT and MEAM results\\
4&interstitial& the formation energy of self-interstitial point defect compared with DFT results\\
5&surface energy& the ($hkl$) surface energy compared with DFT and MEAM results\\
6&phonon bands & the phonon band structures compared with experimental and MEAM results\\
7&stacking-fault energies&the (111) stacking-fault energy compared with DFT and MEAM results\\
\hline
\end{tabular}
\end{table*}

In this section, we report the details of the process that we follow to generate a DP model for Cu with DP-GEN, and demonstrate its uniform accuracy when used to predict a broad range of properties.

\subsection{Generation of the model}
To perform DP-GEN, we used LAMMPS for exploration, VASP~5.4.4 for labeling, and DeePMD-kit for training.
In total, 48 iterations were undertaken. 
Among a total number of 25 million configurations sampled in the exploration step, 7646  (0.03 $\%$) are selected for labeling. 

\textit{Exploration} 
To kick off the DP-GEN procedure, we start with relaxed fcc, hcp, and bcc structures.
The exploration in the first iteration is essentially random.
For each crystalline structure, a  $2 \times 2 \times 2$ supercell is first relaxed and compressed uniformly with a scaling factor $\alpha$ ranging from 0.84 to 1.00.
Then the atomic positions and cell vectors are randomly perturbed. 
The magnitude of perturbations is 0.01 \AA~for the atomic coordinates, and is 3\% of the cell vector length for the simulation cell. 
For each crystal structure, 50 randomly perturbed structures, whose $\alpha$ are $1.00$, and 10 randomly perturbed  structures whose $\alpha$ range from 0.84 to 0.98, are then utilized to perform a 20-step canonical AIMD simulation at $T$=50K with VASP. 

In later exploration steps, we also need to prepare initial 
structures for DPMD sampling.
For bulk systems, we choose the perturbed structures whose $\alpha$ are  $1.00$ for the  three crystal structures.  
For surface systems, we rigidly displace two crystalline halves along crystallographic directions (100), (110), (111) for fcc, and (100), (001), (110) for hcp structures. 
The magnitude of the vacuum slab thickness ranges from 0.5~\AA $ $  to 9~\AA.
The strategy for the bulk system adopts the protocol introduced in Sec.~\ref{sec:method}. 
For the surface systems, there are  four iterations in each temperature interval, and the canonical ensemble (NVT) is used to explore the configurations.
More details of the exploration strategy can be found in Table \ref{table:protocal}. 
The directories of the initial configurations and the parameters (e.g. ensemble, length of the trajectories, sampling frequency) of the MD simulations in each iteration are specified by \texttt{sys$\_$configs} and \texttt{model$\_$devi$\_$jobs}, respectively.    
Another vital component for the selection of configurations is the trust levels.  
We set $\sigma_{lo}$ and $\sigma_{hi}$ to 0.05 eV/\AA~ and 0.2~eV/\AA, respectively:
\begin{verbatim}
    "model_devi_f_trust_lo":	0.05,
    "model_devi_f_trust_hi":	0.20,
\end{verbatim}
      
\textit{Labeling} The Perdew-Burke-Ernzerhof (PBE) generalized gradient approximation~\cite{Perdew1996PBE} is utilized.
The kinetic energy cut-off for the plane wave~\cite{kresse1996efficient, kresse1996efficiency} is set to 650 eV, and the K-points is set using the Monkhorst-Pack mesh~\cite{monkhorst1976special} with the spacing $h_k = 0.1$ \AA$^{-1}$. 
The self-consistent-field iteration will stop when the total energy and band structure energy differences between two consecutive steps are smaller than $10^{-6}$ eV.
If the number of candidate configurations of a crystal structure is larger than 300 for a bulk system, they are randomly down-sampled to 300 before labeling in each iteration. 
For surface systems,  the maximal number of candidates to be labeled is 100 for each structure. 
This can be controlled by the key \texttt{fp\_task\_max}. 

Users can also conveniently designate the directories of general settings (\texttt{INCAR}) and  pseudo-potentials (\texttt{POTCAR}) in \texttt{PARAM}. 
\begin{verbatim}
    "fp_pp_path":   "/path/to/POTCAR",
    "fp_incar":     "/path/to/INCAR"
\end{verbatim}

After the DP-GEN workflow finishes, a representative set of training data is generated.
To properly obtain the energy of a single copper atom in the vacuum, we perform an additional DFT calculation for a single copper atom located in a box of 19 \AA.
The corresponding data is duplicated 200 times and added to the training dataset.
Finally, a productive DP model is trained with 8,000,000 steps, with the learning rates exponentially decaying from $ 1.0  \times 10^{-3} $ to $3.5 \times 10^{-8} $.

\textit{Training}. 
In this work, the smooth version of the DP model is adopted~\cite{zhang2018end}. 
The cut-off radius is set to 8~\AA, and the inverse distance $1/r$ decays smoothly from $2$~\AA\ to 8~\AA\ in order to remove the discontinuity introduced by the cut-off.  
The embedding network of size (25, 50, 100) follows a ResNet-like architecture~\cite{He_2016_CVPR}.  
The fitting network is composed of three layers, with each containing 240 nodes. 
The Adam stochastic gradient descent method~\cite{Kingma2015adam} is utilized to train four models, with the only difference being their random seeds. 
Each model is trained with 400,000 gradient descent steps with an exponentially decaying
 learning rate from $ 1.0  \times 10^{-3} $ to $3.5 \times 10^{-8} $. 
These settings for the training process are designated by the key \texttt{default\_training\_params} provided by the input file PARAM, which adopts the same interface with DeePMD-kit.

\textit{Machine settings}. For training and exploration, we use single card of Tesla-P100 for each calculation task. 
Labeling is done on  Intel(R) Xeon(R) Gold 5117 CPU @ 2.00GHz with 28 CPU cores. The high performance cluster is  scheduled by Slurm. 
    
In MACHINE file, one can specify the settings according to his/her own environment. We provide an easy example for the training machine settings on a Slurm system:
\begin{verbatim}
"machine": {
    "batch":        "slurm",
    "hostname":     "localhost",
    "port":         22,
    "username":     "user",
    "work_path":    "/path/to/Cu/work"
},
"resources": {
    "numb_node":    1,
    "numb_gpu":     1,
    "task_per_node":4,
    "partition":    "GPU",
    "source_list":  ["/path/to/env"]
}
\end{verbatim}
    
Settings in \texttt{machine} specify how to connect with the computational machine and how to transfer files. 
All tasks will be sent to and run in the directories {work$\_$path} on the computational machines.
Reading the keyword \texttt{resources}, DP-GEN  automatically requires for resources that users need among the machines, such as the number of GPU cards and CPU cores, the prerequisite modules for the software that performs the calculation, etc.

\subsection{Testing the model}

To validate the performance of the DP model for pure copper, we test it on a wide range of properties, as summarized in Table \ref{tab:TestDesign}. 
We compare the results of the DP model in predicting the important material properties with a state-of-the-art empirical FF model, obtained by the modified embedded atom method (MEAM)~\cite{baskes1992modified}.

The equilibrium properties of Cu are presented in
Table~\ref{tab:result}. These include the atomization energy and equilibrium volume per atom, defect formation energies, elastic constants and moduli, and stacking-fault energies. The defect formation energy is defined as $E_{\text{df}} = E_{\text{d}}(N_{\text{d}})-N_{\text{d}}E_0$. 
Here, $d = v(i)$ indicates vacancy (self-interstitial) defects, $E_{\text{d}}(N_{\text{d}})$ denotes the relaxed energy of a defective structure with $N_{\text{d}}$ atoms and $E_0$ denotes the energy per atom of the corresponding ideal crystal at $T$ = 0 K.  
Besides, we replicate the primitive fcc cell $3\times3\times3$ times to generate a supercell, and use it to compute the defect formation energies. For all the properties listed in Table \ref{tab:result}, the DP predictions agree well with DFT and/or experiments.
 
\begin{table}[htbp]
\centering
\caption{\label{tab:result}Equilibrium properties of Cu: atomization energy $E_{\text{am}}$,
equilibrium volume per atom $V_0$, vacancy formation energy $E_{\text{vf}}$,
self-interstitial point formation energies $E_{\text{if}}$ for octahedral interstitial (oh) and
tetrahedral interstitial (th) , independent elastic constants $C_{11}$, $C_{12}$,
and $C_{44}$, bulk modulus $B_{\text{V}}$ (Voigt), shear modulus $G_{\text{V}}$ (Voigt), stacking fault energy $\gamma_{\text{sf}}$.}
\begin{tabular}{lcccc}
\hline
\hline
Cu&EXP&DFT\footnote{The DFT results are computed by
the authors. While a smaller K-mesh spacing in DFT may lead to more converged results,  we set the  K-mesh spacing equal to 0.1 $\text{\r{A}}^{-1}$, in order to be consistent with settings in the labeling step. }&DP\footnote{The numbers in parentheses are
standard deviations among 4 models in the last one digit}&MEAM\\
\hline
$E_{\text{am}}$(eV/atom)&-3.563\footnote{Reference\cite{medvedev1989codata}}&-3.712&-3.7098$(1)$&-3.540\\
$V_0$($\text{\r{A}}^{3}$/atom)\footnote{Experiment value was extrapolated to absolute zero and corrected for zero-point vibrations; DFT, DP and MEAM results obtained at $T=0$K. }&11.65\footnote{Reference\cite{lejaeghere2014error}}&12.00&12.004$(1)$&11.65\\
$E_{\text{vf}}$(eV)&1.29\footnote{Reference\cite{triftshauser1975positron}}&1.020&0.99$(1)$&1.105\\
$E_{\text{if}}$(oh)(eV)&&3.616&3.472$(6)$&3.136\\
$E_{\text{if}}$(th)(eV)&&3.999&4.149$(7)$&4.604\\
$C_{11}$(GPa)&176.2\footnote{Reference\cite{overton1955temperature}\label{elas}}&171.74&173$(2)$&175.76\\
$C_{12}$(GPa)&124.9\textsuperscript{\ref{elas}}&118.91&122$(2)$&124.09\\
$C_{44}$(GPa)&81.77\textsuperscript{\ref{elas}}&81.59&75.7$(3)$&77.59\\
$B_{\text{V}}$(GPa)&142.0\textsuperscript{\ref{elas}}&136.52&139$(1)$&141.31\\
$G_{\text{V}}$(GPa)&59.32\textsuperscript{\ref{elas}}&58.32&55.8$(5)$&56.89\\
$\gamma_{\text{sf}}(\text{mJ}/\text{m}^{2})$&41\footnote{Reference\cite{stobbs1971weak}}&38.08&36$(2)$&72.7\\
\hline
\hline
\end{tabular}
\end{table}
 
\begin{figure}[htbp]
\includegraphics[width=9cm]{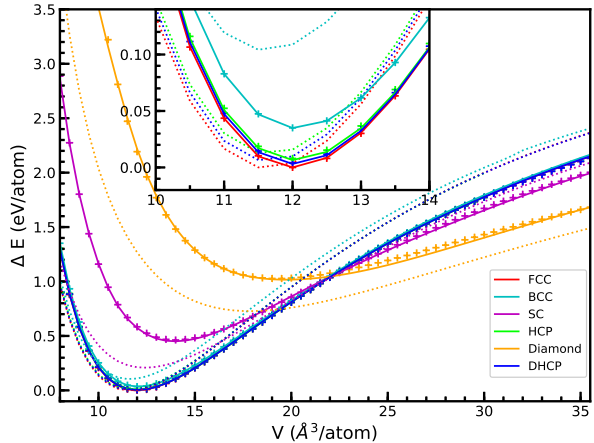}
\centering
\caption{EOS curves for Cu. Solid lines, dotted lines, and plus markers
denote DP, MEAM, and DFT results, respectively. The energies of
MEAM are shifted so that the MEAM energy of a stable fcc structure
equals that given by DFT. 
DFT based relaxations fail for some hcp structures with volume per atom  larger than 30 $\text{\r{A}}^{3}$. Therefore, the corresponding DFT predictions are not shown.  
The diamond, sc, and dhcp structures are not explicitly included in the training data of the DP model.}

\label{fig:eos}
\end{figure}

\begin{figure}[htbp]
\includegraphics[width=9cm]{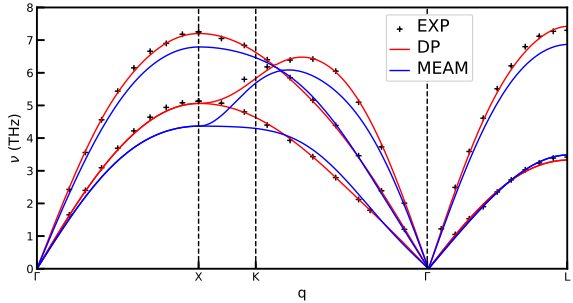}
\centering
\caption{
The DP and MEAM results of phonon dispersion relations for Cu are calculated by phonopy~\cite{phonopy} and its LAMMPS interface  phonolammps~\cite{phonolammps} at $T$ = 49 K with the supercell size being $10\times10\times10$.  Here $q$ denotes the wave number and $\nu$ the frequency. The experimental data  are taken from~Ref.\cite{nicklow1967phonon}. }
\label{fig:phonon}
\end{figure}

The predictions via DFT, DP, and MEAM for the equation of state (EOS) are presented
 in Fig.\ref{fig:eos}. 
DP reproduces well the DFT results for all the standard crystalline structures considered here, i.e., fcc, hcp, double hexagonal close-packed (dhcp), bcc, simple cubic (sc) and diamond. 
It is worth noting that the diamond, sc and dhcp structures are not explicitly explored by the DP-GEN scheme, i.e., the initial training data and the initial structures for exploration do not contain these crystal structures.
Nevertheless, DP  still achieves a satisfactory accuracy in the EOS test on these three structures. 
In comparison,  although MEAM performs well for fcc, hcp and dhcp structures near the energy minimum, it shows large deviations when predicting the EOS for sc, diamond and bcc structures. 
We also report the DP and MEAM predictions for the phonon dispersion relations as well as experimental results.
As shown in Fig. \ref{fig:phonon}, DP results agree very well with the experiment and are significantly better than MEAM predictions.
 
\begin{figure}[htbp]
\includegraphics[width=9cm]{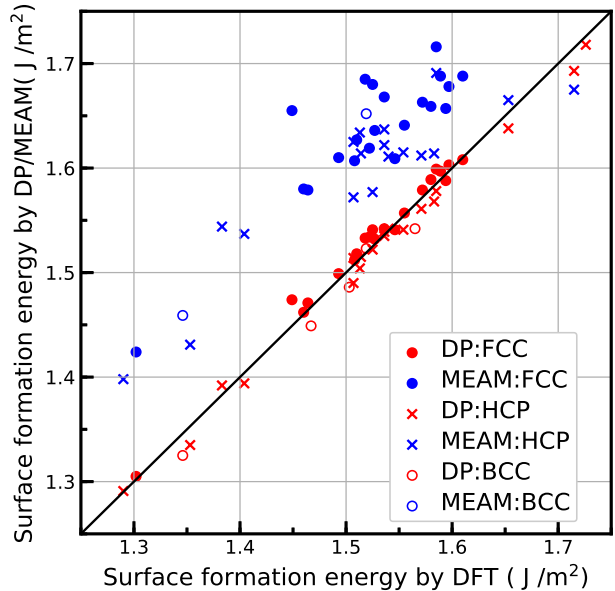}
\centering
\caption{Surface formation energies for fcc, hcp and bcc-lattices of Cu. All the
nonequivalent surfaces with Miller index values smaller than 4
are taken into the consideration for the fcc lattice. 
For the hcp and bcc lattices, the results for the nonequivalent surfaces with Miller index values smaller than 3  are included.} 
\label{fig:surf}
\end{figure}

Finally, we consider the surface formation energy
$E_{\text{sf}}((hkl))$, which describes the energy needed to create a
surface with  Miller indices ($hkl$) for a given crystal, and is defined by $E_{\text{sf}}((hkl)) = \frac{1}{2A} [E_{\text{s}}((hkl)) -N_{\text{s}}E_0]$. Here $E_{\text{s}}((hkl))$
and $N_{\text{s}}$ denote the energy and number of atoms of the relaxed surface structure with Miller indices ($hkl$). 
$A$ denotes the surface area. 
We enumerate all the nonequivalent surfaces with the Miller indices smaller than 4 for the fcc lattice and smaller than 3 for the hcp and bcc lattices. 
As shown in Fig. \ref{fig:surf}, despite the lack of any explicitly labeled data for bcc surfaces in the training dataset,  the surface formation energies predicted by DP are close to those by  DFT, and are significantly better than those predicted by MEAM.

\section{Conclusion}
In this paper, we introduced the software platform DP-GEN. 
We described its implementation and reported the details when used to generate a general purpose PES model for Cu.  
We expect DP-GEN to be a scalable and flexible platform. 
The three steps, exploration, training, and labeling, which are controlled by the scheduler, are separate and highly modularized. 
Therefore, developers will spend a minimal amount of effort to incorporate novel functionalities. 
For example, DP-GEN can easily be extended to include a different first-principles code, 
for which typically only file conversion and job submission scripts are needed.
Furthermore, a list of sampling techniques may be added to the exploration step.

Provided the ability of the DP model to describe various systems, such as organic molecules, metals, semiconductors, and insulators, we expect DP-GEN to be widely used in different molecular and  materials science applications.
Such applications are not limited to the generation of general purpose PES models, but also include the investigation of specific problems.

The DP-GEN workflow can be made applicable to different software and operating systems.
In particular, it can be easily implemented on popular cloud machines such as the Amazon Web Services (AWS) cloud platform~\cite{aws}.
Moreover, there have been significant efforts to build automatic interactive platforms for computational science.
Among these efforts, the AiiDA package~\cite{pizzi2016aiida} has become very promising for creating social ecosystems to disseminate codes, data, and scientific workflows.
To connect DP-GEN with popular general-purpose open-source platforms is on our to-do list.
Above all, we expect that users of DP-GEN will be embraced with an optimal solution in terms of both efficiency and cost when performing atomic and molecular simulations. 

\section{Acknowledgement}
The authors thank Marcos F. Calegari Andrade, Hsin-Yu Ko, Jianxing Huang, Yunpei Liu, Mengchao Shi, Fengbo Yuan, and Yongbin Zhuang for helps and discussions. 
We are grateful for computing time provided by the TIGRESS High Performance Computer Center at Princeton University, the High-performance Computing Platform of Peking University, and the Beijing Institute of Big Data Research.
The work of L. Z. and W. E was supported in part by a gift from iFlytek to Princeton University, the ONR grant N00014-13-1-0338, and the Center Chemistry in Solution and at Interfaces (CSI) funded by the DOE Award DE-SC001934.
The work of Han Wang is supported by the National Science Foundation of China under Grant No. 11871110, the National Key Research and Development Program of China under Grants No. 2016YFB0201200 and No. 2016YFB0201203, and Beijing Academy of Artificial Intelligence (BAAI). 
The work of J. Z. is partially supported by National Innovation and Entrepreneurship Training Program for Undergraduate (201910269080).

\bibliographystyle{apsrev4-1}
\bibliography{references}

\newpage
\appendix
\setcounter{table}{0}
\setcounter{figure}{0}
\renewcommand\thetable{\Alph{section}.\arabic{table}}

\section{Details of the exploration strategy and more numerical results}
\begin{table*}[htbp]
\tiny
\begin{tabular}{ccccccc}
\hline
Iter. & Crystal & \#DPMD &   Length (ps) & T (K) & Ensemble & Candidate Per($\%$) \\ 
\hline
0 & FCC, HCP, BCC& 600 & 2 & 50,135,271,407,543& NPT & 8.29 \\
1 & FCC, HCP, BCC& 600 & 2 & 50,135,271,407,543& NPT & 0.00 \\
2 & FCC, HCP, BCC& 1200 & 6 & 50,135,271,407,543& NPT & 0.00 \\
3 & FCC, HCP, BCC& 1200 & 6 & 50,135,271,407,543& NPT & 0.00 \\
4 & FCC, HCP, BCC& 1200 & 6 & 50,135,271,407,543& NPT & 0.00  \\
5 & FCC, HCP, BCC& 1200 & 6 & 50,135,271,407,543& NPT & 0.00 \\
6 & FCC, HCP, BCC& 2400 & 6 &  50,135,271,407,543& NPT & 0.00  \\
7 & FCC, HCP, BCC& 2400 & 6 & 50,135,271,407,543&  NPT & 0.00 \\
8 & FCC, HCP, BCC& 600 & 2 & 678,814,950,1086,1221&  NPT &  5.45 \\
9 & FCC, HCP, BCC& 600 & 2 & 678,814,950,1086,1221&  NPT &   0.00 \\
10 & FCC, HCP, BCC& 1200 & 6 &  678,814,950,1086,1221&  NPT &  0.01  \\
11 & FCC, HCP, BCC& 1200 & 6 &  678,814,950,1086,1221&  NPT &  0.01 \\
12 & FCC, HCP, BCC& 1200 & 6 &  678,814,950,1086,1221&  NPT &  0.01  \\
13 & FCC, HCP, BCC& 1200 & 6 &  678,814,950,1086,1221&  NPT &  0.00  \\
14 & FCC, HCP, BCC& 2400 & 6 &  678,814,950,1086,1221&  NPT &  0.00  \\
15 & FCC, HCP, BCC& 2400 & 6 &  678,814,950,1086,1221&  NPT &  0.00  \\
16 & FCC, HCP, BCC& 600 & 2 &  1357,1493,1629,1765,1900&  NPT &   2.96 \\
17 & FCC, HCP, BCC& 600 & 2 &  1357,1493,1629,1765,1900&  NPT &   0.00  \\
18 & FCC, HCP, BCC& 1200 & 6 &  1357,1493,1629,1765,1900&  NPT &   0.01 \\
19 & FCC, HCP, BCC& 1200 & 6 &  1357,1493,1629,1765,1900&  NPT &  0.01  \\
20 & FCC, HCP, BCC& 1200 & 6 &  1357,1493,1629,1765,1900&  NPT &   0.01 \\
21 & FCC, HCP, BCC& 1200 & 6 &  1357,1493,1629,1765,1900&  NPT &   0.01 \\
22 & FCC, HCP, BCC& 2400 & 6 &  1357,1493,1629,1765,1900&  NPT &  0.00  \\
23 & FCC, HCP, BCC& 2400 & 6 &  1357,1493,1629,1765,1900&  NPT &   0.00 \\
24 & FCC, HCP, BCC& 600 & 2 &  2036,2172,2308,2443,2579&  NPT &   0.06 \\
25 & FCC, HCP, BCC& 600 & 2 &  2036,2172,2308,2443,2579&  NPT &   0.04 \\
26 & FCC, HCP, BCC& 1200 & 6 &  2036,2172,2308,2443,2579&  NPT &   0.02 \\
27 & FCC, HCP, BCC& 1200 & 6 &  2036,2172,2308,2443,2579&  NPT &   0.08 \\
28 & FCC, HCP, BCC& 1200 & 6 &  2036,2172,2308,2443,2579&  NPT &   0.00 \\
29 & FCC, HCP, BCC& 1200 & 6 &  2036,2172,2308,2443,2579&  NPT &   0.00 \\
30 & FCC, HCP, BCC& 2400 & 6 &  2036,2172,2308,2443,2579&  NPT &   0.01 \\
31 & FCC, HCP, BCC& 2400 & 6 &  2036,2172,2308,2443,2579&  NPT &   0.00 \\
32 & FCC (Surf), HCP (Surf)& 2400 & 2 &  50,135,271,407,543& NVT & 53.97  \\
33 & FCC (Surf), HCP (Surf)& 2400 & 2 & 50,135,271,407,543& NVT & 0.16  \\
34 & FCC (Surf), HCP (Surf)& 4800 & 6 & 50,135,271,407,543& NVT & 0.04  \\
35 & FCC (Surf), HCP (Surf)& 4800 & 6 & 50,135,271,407,543& NVT & 0.00  \\
36 & FCC (Surf), HCP (Surf)& 2400 & 2 & 678,814,950,1086,1221& NVT & 0.01  \\
37 & FCC (Surf), HCP (Surf)& 2400 & 2 & 678,814,950,1086,1221& NVT & 0.16  \\
38 & FCC (Surf), HCP (Surf)& 4800 & 6 & 678,814,950,1086,1221& NVT & 0.02  \\
39 & FCC (Surf), HCP (Surf)& 4800 & 6 & 678,814,950,1086,1221& NVT & 0.01  \\
40 & FCC (Surf), HCP (Surf)& 2400 & 2 & 1357,1493,1629,1765,1900& NVT  & 0.16 \\
41 & FCC (Surf), HCP (Surf)& 2400 & 2 &  1357,1493,1629,1765,1900& NVT  &  0.05 \\
42 & FCC (Surf), HCP (Surf)& 4800 & 6 &  1357,1493,1629,1765,1900& NVT & 0.22 \\
43 & FCC (Surf), HCP (Surf)& 4800 & 6 &  1357,1493,1629,1765,1900& NVT & 0.04 \\
44 & FCC (Surf), HCP (Surf)& 2400 & 2 &  2036,2172,2308,2443,2579& NVT & 0.05  \\
45 & FCC (Surf), HCP (Surf)& 2400 & 2 &  2036,2172,2308,2443,2579& NVT & 0.08 \\
46 & FCC (Surf), HCP (Surf)& 4800 & 6 &  2036,2172,2308,2443,2579& NVT & 0.03 \\
47 & FCC (Surf), HCP (Surf)& 4800 & 6 &  2036,2172,2308,2443,2579& NVT & 0.24  \\
\hline
\end{tabular}
\caption{\label{table:protocal}Exploration strategy for the copper system. For each iteration, we report the crystalline structure from which the initial structures derive, the number of DPMD simulations, the length of trajectories, the simulation temperatures, the statistical ensembles, and the percentages of candidates for labeling. For those simulations which adopt the NPT ensemble,  1,
  10
  100,
  1,000,
  5,000,
  10,000,
  20,000 and
 50,000~Bar are set as the pressures. }
\end{table*}

\begin{table*}[htbp]
\tiny
\begin{tabular}{ccccccc}
\hline
fcc & Miller indices ($h.k.l$) & $E_{sf}^0$ &$E_{sf}^1$ &$E_{sf}^2$&$E_{sf}^3$& $\sigma(E_{sf})$ \\
\hline

000 & 1.1.1 	&	1.302	&	1.306	&	1.298	&	1.301	&	0.003	\\
001 & 3.3.2 	&	1.488	&	1.492	&	1.496	&	1.498	&	0.004	\\
002 & 3.3.1 	&	1.532	&	1.537	&	1.545	&	1.544	&	0.005	\\
003 & 1.1.0 	&	1.476	&	1.479	&	1.483	&	1.482	&	0.003	\\
004 & 3.3.-1 	&	1.526	&	1.530	&	1.536	&	1.536	&	0.004	\\
005 & 3.3.-2 	&	1.536	&	1.540	&	1.545	&	1.544	&	0.004	\\
006 & 1.1.-1 	&	1.531	&	1.533	&	1.535	&	1.534	&	0.001	\\
007 & 3.2.2 	&	1.505	&	1.511	&	1.512	&	1.513	&	0.003	\\
008 & 3.2.1 	&	1.611	&	1.612	&	1.610	&	1.613	&	0.001	\\
009 & 3.2.-1 	&	1.597	&	1.600	&	1.603	&	1.605	&	0.003	\\
010 & 3.2.-2 	&	1.598	&	1.600	&	1.600	&	1.602	&	0.001	\\
011 & 3.2.-3 	&	1.581	&	1.581	&	1.580	&	1.584	&	0.002	\\
012 & 3.1.-1 	&	1.591	&	1.592	&	1.591	&	1.594	&	0.001	\\
013 & 3.1.-2 	&	1.607	&	1.608	&	1.605	&	1.610	&	0.002	\\
014 & 3.1.-3 	&	1.591	&	1.592	&	1.590	&	1.594	&	0.001	\\
015 & 3.0.-1 	&	1.459	&	1.466	&	1.463	&	1.470	&	0.004	\\
016 & 3.0.-2 	&	1.531	&	1.535	&	1.534	&	1.536	&	0.002	\\
017 & 1.0.-1 	&	1.545	&	1.545	&	1.544	&	1.544	&	0.000	\\
018 & 3.-1.-1 	&	1.450	&	1.455	&	1.454	&	1.459	&	0.003	\\
019 & 3.-1.-2 	&	1.553	&	1.555	&	1.556	&	1.559	&	0.002	\\
020 & 3.-2.-2 	&	1.511	&	1.514	&	1.517	&	1.518	&	0.003	\\
\hline
\hline
hcp & Miller indices ($h.k.l$) & $E_{sf}^0$ &$E_{sf}^1$ &$E_{sf}^2$&$E_{sf}^3$& $\sigma(E_{sf})$ \\
\hline
000 & 1.1.1	&	1.688	&	1.694	&	1.700	&	1.698	&	0.005	\\
001 & 1.1.1	&	1.387	&	1.388	&	1.388	&	1.386	&	0.001	\\
002 & 2.2.1	&	1.389	&	1.392	&	1.395	&	1.397	&	0.003	\\
003 & 2.2.1	&	1.573	&	1.576	&	1.582	&	1.582	&	0.004	\\
004 & 1.1.0	&	1.713	&	1.714	&	1.725	&	1.721	&	0.005	\\
005 & 1.1.0	&	1.329	&	1.327	&	1.328	&	1.327	&	0.001	\\
006 & 2.1.2	&	1.510	&	1.509	&	1.513	&	1.510	&	0.001	\\
007 & 2.1.1	&	1.562	&	1.560	&	1.563	&	1.561	&	0.001	\\
008 & 2.1.0	&	1.517	&	1.515	&	1.518	&	1.517	&	0.001	\\
009 & 2.-1.2	&	1.534	&	1.532	&	1.538	&	1.537	&	0.002	\\
010 & 2.-1.2	&	1.530	&	1.531	&	1.539	&	1.538	&	0.004	\\
011 & 2.-1.1	&	1.535	&	1.536	&	1.541	&	1.542	&	0.003	\\
012 & 2.-1.1	&	1.536	&	1.535	&	1.540	&	1.539	&	0.002	\\
013 & 2.-1.0	&	1.485	&	1.488	&	1.489	&	1.492	&	0.002	\\
014 & 2.-1.0	&	1.633	&	1.630	&	1.631	&	1.633	&	0.001	\\
015 & 2.-2.1	&	1.556	&	1.555	&	1.556	&	1.556	&	0.000	\\
016 & 1.1.2	&	1.508	&	1.508	&	1.517	&	1.514	&	0.004	\\
017 & 1.1.2	&	1.499	&	1.504	&	1.509	&	1.511	&	0.005	\\
018 & 0.0.1	&	1.285	&	1.285	&	1.278	&	1.281	&	0.003	\\
\hline
\hline
bcc & Miller indices ($h.k.l$) & $E_{sf}^0$ &$E_{sf}^1$ &$E_{sf}^2$&$E_{sf}^3$& $\sigma(E_{sf})$ \\
\hline
000 & 1.1.1	&	1.541	&	1.537	&	1.544	&	1.542	&	0.003	\\
001 & 2.2.1	&	1.484	&	1.486	&	1.494	&	1.493	&	0.004	\\
002 & 1.1.0	&	1.323	&	1.333	&	1.332	&	1.327	&	0.004	\\
003 & 2.1.1	&	1.214	&	1.208	&	1.214	&	1.212	&	0.002	\\
004 & 2.1.0	&	1.447	&	1.45	&	1.459	&	1.457	&	0.005	\\
005 & 1.0.0	&	1.521	&	1.525	&	1.535	&	1.53	&	0.005 \\
\hline
\end{tabular}
\caption{\label{table:surface_std} The predictions of surface formation energies $E_{sf}$ for fcc, hcp and bcc-lattices and their standard deviations $\sigma(E_{sf})$ among 4 models.}
\end{table*}

\end{document}